\documentclass[sigconf]{acmart}

\copyrightyear{2021}
\acmYear{2021}
\setcopyright{acmcopyright}
\acmConference[HT '21] {Proceedings of the 32nd ACM Conference on Hypertext and Social Media}{August 30--September 2, 2021}{Virtual Event, Ireland.}
\acmBooktitle{Proceedings of the 32nd ACM Conference on Hypertext and Social Media (HT '21), August 30--September 2, 2021, Virtual Event, Ireland}
\acmPrice{15.00}
\acmDOI{10.1145/3465336.3475097}
\acmISBN{978-1-4503-8551-0/21/08}

\usepackage{url}
\usepackage[utf8]{inputenc}
\usepackage{graphicx}
\usepackage{amsmath}
\usepackage{amsthm}
\usepackage{booktabs}
\usepackage{algorithm}
\usepackage{algorithmic}
\usepackage{multirow}
\usepackage{caption}
\usepackage{subcaption}
\usepackage{balance}
\AtBeginDocument{%
  \providecommand\BibTeX{{%
    \normalfont B\kern-0.5em{\scshape i\kern-0.25em b}\kern-0.8em\TeX}}}

\setcopyright{none}
\begin{document}
\fancyhead{}

\title{Profiling Fake News Spreaders on Social Media \\ through Psychological and Motivational Factors}

\author{Mansooreh Karami}
\affiliation{%
  \institution{Arizona State University}
  \city{Tempe}
  \state{AZ}
  \country{USA}
}
\email{mkarami@asu.edu}

\author{Tahora H. Nazer}
\affiliation{%
  \institution{Spotify}
  \city{Boston}
  \state{MA}
  \country{USA}}
\email{tahoranazer@spotify.com}

\author{Huan Liu}
\affiliation{%
  \institution{Arizona State University}
  \city{Tempe}
  \state{AZ}
  \country{USA}
}
\email{huanliu@asu.edu}









\renewcommand{\shortauthors}{Trovato and Tobin, et al.}

\begin{abstract}
  

The rise of fake news in the past decade has brought with it a host of consequences, from swaying opinions on elections to generating uncertainty during a pandemic.
 A majority of methods developed to combat disinformation either focus on fake news content or malicious actors who generate it.
However,  the virality of fake news is largely dependent upon the users who propagate it.
A deeper understanding of these users can contribute to the development of a framework for identifying users who are likely to spread fake news.
  In this work, we study the characteristics and motivational factors of fake news spreaders on social media with input from psychological theories and behavioral studies.
We then perform a series of experiments to determine if fake news spreaders can be found to exhibit different characteristics than other users.
Further, we investigate our findings by testing whether the characteristics we observe amongst fake news spreaders in our experiments can be applied to the detection of fake news spreaders in a real social media environment.

\end{abstract}




\maketitle

\section{Introduction}

Due to the increasing amount of our time spent on social media platforms, it is no surprise that people tend to receive their news content through social media more than before. One in five U.S. adults used social media as their main source of political and election news for the US presidential election in 2020~\cite{amymitchell2020}.
The high rate of engagement with online news can be mainly attributed to the nature of the social media platforms themselves. Social Media is typically inexpensive, provides easy access to users, and supports fast dissemination of information that is not possible through traditional media outlets. However, despite these advantages, the quality of news on social media is considered lower than that of traditional news outlets. A factor contributing to this low quality is the widespread nature of fake news articles online. Fake news is a piece of false information published by news outlets to mislead consumers~\cite{zhou2020survey, shu2020combating}.


Fake news has several significant negative effects on civil society. First, people may accept deliberate lies as truths. The likelihood of accepting fake news as true increases after repeated exposure~\cite{hasher1977frequency}, especially when the content aligns with the user's beliefs~\cite{weir2017believe, jiang2021mechanisms}. Second, fake news may change the way people respond to legitimate news. When people are inundated with fake news, the line between fake news and real news becomes more uncertain. Fake news spreaders make users doubt the nature of real news and create the idea that everything is biased and conflicted, and it is impossible to distinguish fake from real news~\cite{Lynch2016fake}. Finally, the prevalence of fake news has the potential to break the trustworthiness of the entire news ecosystem. 
For instance, despite traditional domains such as the New York Times, the Washington Post, and CNN being among the most shared COVID-19-related stories on Twitter, a fake news domain, Gateway Pundit, was ranked 4th in August and 6th in September of 2020 among the most shared domains for URLs about COVID-19~\cite{lazer2020state}.
Therefore, it is critical to develop methods that detect and mitigate fake news, with the purpose of benefiting the general public and the entire news ecosystem.


Detecting fake news is a challenging task because it is designed to be indistinguishable from real news and intentionally misleading. As a result, the features extracted from the content are not enough to build an accurate detection method. Researchers have tackled this problem by using the characteristics of posts, user networks, and user profiles. However, factors that motivate users to spread fake news on social media and their importance in fake news detection have remained understudied. Studies~\cite{pennycook2020fighting,difonzo2007rumor} have shown that users are more prone to propagating fake news stories when the situation is uncertain, they are emotionally overwhelmed and anxious, the topic of discussion is of personal importance to them, and they do not have primary control over the situation through their actions. In this work, we provide a methodology to extract factors that are not reliant on currently existing self-reported survey methods~\cite{buchanan2020people, talwar2019people}. Our analysis aims to shed light on the attitude of spreading fake news, rather than detecting potential fake news among the users’ tweets. 
We investigate the following research questions:

\noindent
\textbf{RQ1}: \textit{What motivates users to spread fake news?}

\noindent
\textbf{RQ2}: \textit{What are the differences between users who spread fake news and the users who spread real news in terms of motivational factors determined in \textbf{RQ1}?}

\noindent
\textbf{RQ3}: \textit{Can we use the differences between fake news spreaders and other users to profile fake news spreaders?}

We make the following contributions in this work:
\begin{itemize}
  \item We study a novel problem by investigating the relationship between motivational factors and spreading fake news on social media in Sections~\ref{sec:rq1} and~\ref{sec::data}. This study lays the foundation of exploiting the effectiveness of these factors for preventing the dissemination of fake news;
  \item We perform a statistical comparative analysis between users who spread fake news and other users. We hypothesize that users who are keen to spread disinformation may have different motivating characteristics compared to other users. Our study shows that they are significantly different in terms of motivational factors in Section \ref{subseq::sig};
  \item We conduct experiments on real-world datasets to show whether the extracted features add additional value to the representation of the users' tweets in Section \ref{sebseq::classifier}.
  \end{itemize}

\section{Motivational Factors in Spreading Fake News}
\label{sec:rq1}

To investigate the factors that motivate users to spread fake news on social media (\textbf{RQ1}), we begin with enumerating a list of motivational factors in spreading fake news: \textit{uncertainty}, \textit{anxiety}, \textit{lack of control}, \textit{relationship enhancement}, and \textit{rank}.

\subsection{Uncertainty}
Spreading fake news can be a sense-making activity in ambiguous situations. The frequency of fake news sharing increases in uncertain situations such as global pandemics, when the details of the spread are unknown, or political events like an election when people are unsure of the results~\cite{pennycook2020fighting,difonzo2007rumor}. When a crisis happens, people first seek information from credible sources~\cite{crediblecerc}. If there is no such information, people tend to form unofficial social networks to make predictions based on their own judgment and fill the information gap~\cite{rosnow1976rumor}. This might result in the generation of fake news such as injecting one's self with bleach or consumption of highly-concentrated alcohol to kill the SARS-CoV-2 novel coronavirus~\cite{WHO}.
As the uncertainty increases, the reliance on firm beliefs and the unity among the users with the same ideology or in the same group reduces. Hence, users are more prone to accept new information, even false, as a compromise to resolve the uncertain uncertainty. Uncertainty can cause emotions like anxiety and anger which can affect the spread of fake news in other ways~\cite{marcus2017affective, weeks2015emotions}.

\subsection{Emotions} 
Emotional pressure can play an important role in spreading fake news and can be triggered by emotions like anxiety. Anxiety can make people prone to spreading unproven and less accurate claims when transmitting  information~\cite{difonzo2007rumor}. In high anxiety situations, fake news can serve to justify the user's feelings and relieve emotional tension~\cite{allport1947psychology}. Fake news might be used as a method of expressing emotions in anxious situations that allows people to talk about their concerns and receive feedback informally. This results in sense-making and problem-solving~\cite{waddington2012gossip}. For example, during the devastating Hurricane Harvey, 2017, a fake news story accusing Black Lives Matter supporters of blocking first responders from reaching the affected area was spread by more than one million Facebook users~\cite{Grenoble2017hurricane}. Believing and spreading such fake news stories may help the people in disaster areas cope with the anxiety caused by delays in relief efforts~\cite{fernandez2017waiting}. Anxiety can reduce the effects of motivated reasoning~\cite{weeks2015emotions}. Users with higher traits of anxiety pay closer attention to the contemporary information even when it is against their ideology or bias~\cite{mackuen2010civic}.



\subsection{Lack of control}
Fake news represents a method for coping with uncertain and uncontrollable situations. When people are uncertain about their situation, they tend to experience the feeling of lack of control~\cite{bordia2004uncertainty}, which leads to anxiety. When people do not have primary control over their situation
, they resort to secondary control strategies consisting of emotional responses such as predicting the worst to avoid disappointment or attributing events to chance. Some additional secondary control themes are interpreting the meaning of the event and predicting future events~\cite{walker1991virulence}. Studies show that fake news can be used as a way of regaining control over uncertain and uncontrollable circumstances. An example of a situation that temporarily induces the feeling of losing control and the subsequent attempt to regain control is when individuals enter a new environment. In such a setting, individuals engage in activities to manifest their control, boost their performance, and increase satisfaction in order to engage more with their environment~\cite{ashford1996proactivity}. The desire to actively engage with the environment can make social media newcomers more vulnerable to fake news because they are less familiar with the environment and more eager to gain the attention of their network.

\begin{figure*}[ht]
  \centering
  \includegraphics[width=0.8\linewidth]{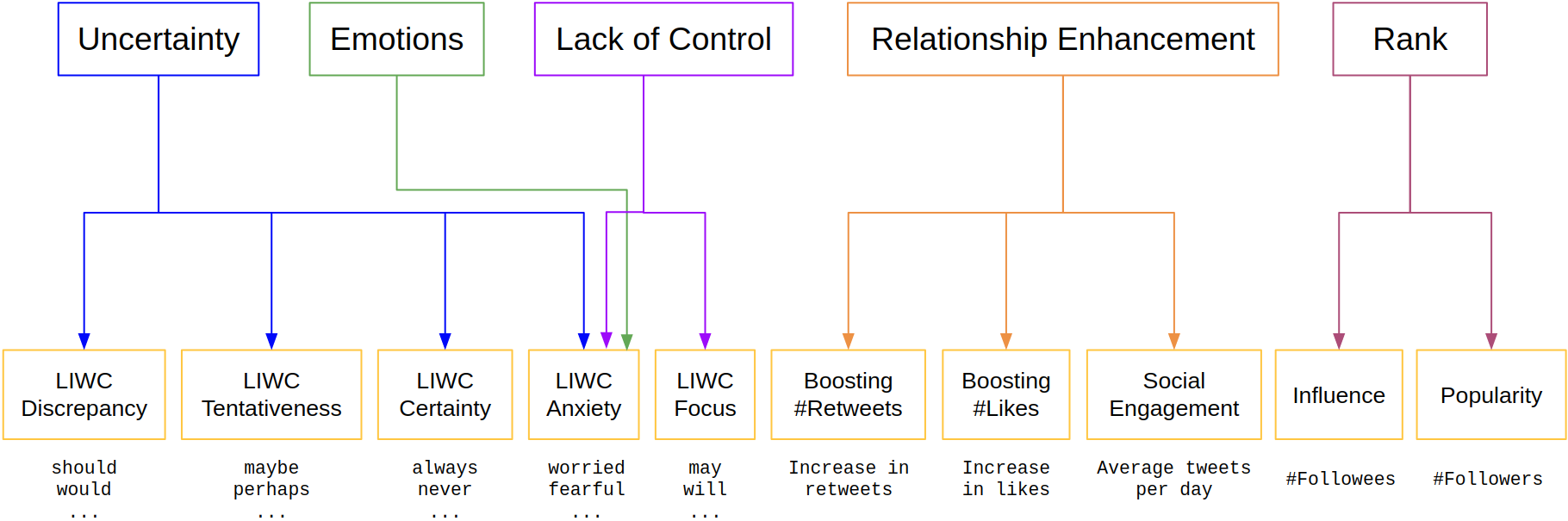}
  \caption{Summary of the metrics used to measure features of users who spread fake news with sample words used.} 
  \label{fig:features} 
\end{figure*}
\subsection{Relationship Enhancement} 
Enhancing social relations is one of the goals of spreading unverified information~\cite{difonzo2007rumor} and conspiracies~\cite{Kou2017conspiracy}. In the formation of short term social relations with the purpose of generating a positive effect on others, people may spread information to attract or hold attention without verifying its authenticity. People prefer to follow, engage with, and endorse those that they like. This liking is caused by three factors: (1)~people that are similar to us, (2)~people that pay us compliments, and (3)~people that cooperate with us towards mutual goals~\cite{cialdini2007influence}. All of these factors can be observed in social cyber communities and cyber tribes. In other words, users on social media tend to share and like news more when it comes from the sources or the communities they follow~\cite{OELDORFHIRSCH2015240}. Also, in friendly relations, negative fake news might be transmitted as a warning mechanism to provide information that is relevant, useful, and prevents harmful consequences~\cite{weenig2001bad}. Essentially, group members want to show that they are looking out for one another. 

\subsection{Rank}
Researchers observed that employees with the lowest rank have a high contribution to the circulation of information that is confirmed to be false~\cite{mitra2012have}. Moreover, spreading fake news can be considered as a self-enhancing mechanism. Self-enhancement motivates individuals to either consciously spread the content for personal gain or unconsciously spread the content that is consistent with their previous beliefs and biases~\cite{buchanan2020people,difonzo2007rumor}. Therefore, the motivation to improve social rank can encourage users to spread fake news on social media. There is not a specific way of defining a rank for users on social media. 
Measures such as popularity or influence are attempts of indicating ranks on social networks.

\section{Data Preparation for Studying Fake News Spreaders}
\label{sec::data}
We use two datasets from FakeNewsNet repository~\cite{shu2020fakenewsnet} to explain our method and experiments in Section \ref{sec:rq2}: PolitiFact and GossipCop\footnote{The source code for this work is available at \href{https://github.com/mansourehk/profiling}{https://github.com/mansourehk/profiling}.}.
The data consists of fake and real stories identified by PolitiFact and GossipCop websites. To identify the users, we utilized a crawler to collect all the tweets referring to those stories. Recent posts, profile, and list of followers and friends of Twitter users who posted about fake and real stories were also collected. We treat each user as a document and the users' recent posts are considered as the content of that document. 
\begin{table}[h]
\centering
\small
\caption{Statistics of the datasets.}
\label{tab:stat}
\begin{tabular}{lccc} 
\toprule
\textbf{Dataset}            & \textbf{News Spreader} & \textbf{\# of Users} & \textbf{\# of Tweets}  \\ 
\hline
\multirow{2}{*}{PolitiFact} & Real                   & 12,511               & 88,930                 \\
                            & Fake                   & 6,309                & 45,184                 \\ 
\hline
\multirow{2}{*}{GossipCop} & Real                   & 2,794                & 134,055                \\
                            & Fake                   & 11,560               & 267,803                \\
\bottomrule
\end{tabular}
\end{table}
We based this method on the experiments done by Yang et al.~\cite{yang2016hierarchical}, using the same parameter of about 150 words per document. Since each tweet consists of at most 280 characters (which on average is about 55 words), if there is a mention or retweet of three or more fake news, that user is tagged as a user who is keen to spread fake news, otherwise, that user is labeled as real news spreader.

All the features introduced in Section \ref{sec:rq1} can be extracted from the content (tweets), activity (tweeting behavior and likes), or network (followees and followers) of Twitter users in PolitiFact and GossipCop datasets. For measuring the features that are extracted from content, we exploit Linguistic Inquiry and Word Count (LIWC)~\cite{tausczik2010psychological}. LIWC includes a dictionary that lists a set of words for some psychologically-relevant categories including positive and negative emotions, social relationships, honesty and deception. 

\subsection{Feature Extraction}
\label{sec:feature_ext}
We pre-processed all tweets to remove punctuation, URLs, hashtags, and mentions. To measure a feature using an LIWC category, we find the percentage of words in a tweet that belongs to that category. Next, we explain how each of the features are measured:
\begin{itemize}
    \item Selected LIWC features: there are three LIWC categories that are related to \textit{uncertainty}: Discrepancy (e.g. should, would, and could), Tentativeness (e.g. maybe, perhaps, and guess), and Certainty (e.g. always and never). These categories are abbreviated as {\tt{discrep}}, {\tt{tentat}}, and {\tt{certain}}, respectively. \textit{Anxiety} can be measured using the LIWC Anxiety category ({\tt{anx}}) which includes words such as nervous, afraid, and tense. 
    We use LIWC Future Focus ({\tt{futurefocus}}) to measure \textit{lack of control}, this category includes words such as may, will, and soon.
    \item Social Engagement: we  measure social engagement on Twitter using the average number of tweets per day.
    \item Position in the Network: this feature can be quantified using a variety of metrics when the network structure is known. However, in the case of social networks between Twitter users, we do not have complete structure and even collecting local information is time consuming due to the rate limitation of Twitter APIs. Hence, we use the information available in our datasets and extract \textit{influence} using the number of followees and \textit{popularity} using the number of followers of each user. 
    \item Relationship Enhancement and Liking: improving the relation to other social media users and gaining more attention from the community is one of the motivations for spreading fake news. If the number of retweets and likes of a user's fake news tweet is higher than the average number of the user's retweets and likes, it can indicate that this user has enhanced their social relation and initiated conversation. Consequently, we use the difference between these two values as indicators of relationship enhancement.
\end{itemize}

The summary of features and the metrics used to measure them is presented in Figure~\ref{fig:features}. As explained in Section \ref{sec:rq1}, some metrics overlap in interpreting the motivational features.

\begin{figure*}[h]
\centering
\begin{subfigure}{.5\textwidth}
  \centering
  \includegraphics[width=1\linewidth]{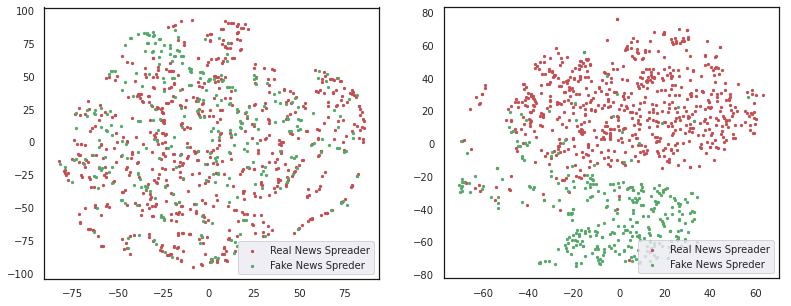}
  \caption{Politifact Dataset}
  \label{fig:sub1}
\end{subfigure}%
\begin{subfigure}{.5\textwidth}
  \centering
  \includegraphics[width=1\linewidth]{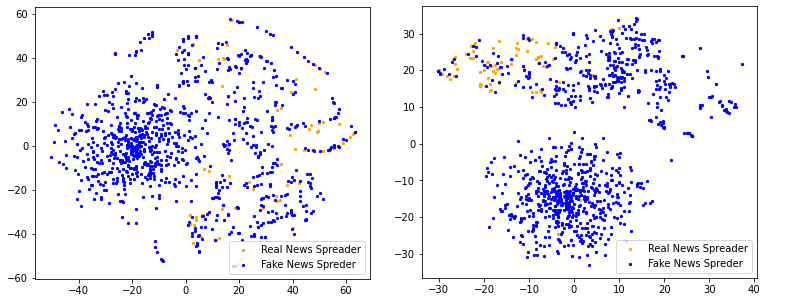}
  \caption{GossipCop Dataset}
  \label{fig:sub2}
\end{subfigure}
\caption{The t-sne visualization of features for the \textsc{Bert} ((a)- and (b)-left) and \textsc{Bert}+Features ((a)- and (b)-right). Adding the motivational features will create a better discriminative embeddings.}
\label{fig:tsne}
\end{figure*}
\section{Profiling Fake News Spreaders}
\label{sec:rq2}
We study the differences between users who spread fake news and the ones who spread real news in terms of the feature categories introduced in Section \ref{sec:rq1} (\textbf{RQ2}). We perform two experiments towards this goal. First, we set our null hypothesis to be that these two groups have the same mean in five categories of features. If the results of t-test shows that there is a significant difference (p-value less than 0.05) between two groups, we can reject the null hypothesis. Second, we use the proposed features to see if a supervised model (a classifier) can use these features to discriminate the two groups of users. The byproduct of this experiment also validates if these features can add more information to the representation of the user's data (\textbf{RQ3}).
\begin{table}[h]
      \centering
      \small
      \captionsetup{margin=0.3cm}
        \begin{tabular}{lcc} 
\toprule
\textbf{Feature~\textbf{Name}} & \textbf{t-statistic~\textbf{Politifact}} & \textbf{t-statistic~\textbf{GossipCop}}  \\
\midrule
Tentativeness                  & -22.91**                                 & 17.80**                                  \\
Discrepancy                    & -16.13**                                 & 16.43**                                  \\
Certainty                      & -8.67**                                  & 16.36**                                  \\
Anxiety                        & -1.06*                                   & 1.02                                     \\
Lack of Control                & -14.66**                                 & 19.18**                                  \\
Social Engagement              & -13.12**                                 & -0.41                                    \\
Influence                      & -0.26                                    & 2.74**                                   \\
Popularity                     & -2.69*                                   & 2.10*                                    \\
Boosting \#tweets              & -3.78**                                  & 9.70**                                   \\
Boosting \#likes               & -1.54                                    & -0.70                                    \\
\bottomrule
\end{tabular}
\caption{Comparison between fake news and real news spreaders in terms of psychological features. The features that are significantly different between users who spread fake news and the users who spread reals news are marked with ** (p-value \textless 0.005) or * (p-value \textless 0.05)}
\label{tab:real_vs_fake_users}
    
\end{table}

\subsection{Significance of Psychological Features}
\label{subseq::sig}

As shown in Table \ref{tab:real_vs_fake_users}, for the features, we observe a significant difference (p-value $<0.05$) between users who share fake news and the ones who share real news in terms of most of the features. 
Notably, the T-test for Anxiety in the GossipCop dataset is not significant. We reason that this is because GossipCop articles are mostly concerned with celebrity news and regardless of the true or false nature of the news being propagated, the news has a low potential of causing distress or being propagated to address the anxiety of the spreader. On the other hand, Politifact contains news articles about a variety of topics including natural disasters and politics. Because these topics are serious, there is more potential for these articles to be circulated due to the anxiety of the spreader.

In social engagement, Politifact users who share fake news have significantly lower number of tweets per day but we do not observe a significant difference among users in the GossipCop users. We used the Influence (number of followees) and popularity (number of followers) as indicators of the status in the network. Except for the Influence in the Politifact users, we observe a significant difference among users who spread real news and the ones spreading fake news. We also expected to see different number of likes and retweets for the fake news tweets which is only observed in reshares feature; indicating that Boosting \#likes can be a motivation for spreading fake news on social media.

\begin{table}[h]

      \centering
      \small
      \captionsetup{margin=0.7cm}
\begin{tabular}{lcc}
    \toprule
        \multicolumn{3}{c}{\bf Politifact Dataset}   \\ \midrule
{\bf Model}   & {\bf Accuracy} & {\bf F1-Score}  \\
\midrule
Bert        &   78.61  &   58.02   \\
Bert+Features   &   90.0  &   83.11  \\
\bottomrule
\multicolumn{3}{c}{\bf GossipCop Dataset}   \\ 
\midrule
{\bf Model}   & {\bf Accuracy} & {\bf F1-Score}  \\
\midrule
Bert        &   90.36  &   94.61   \\
Bert+Features   &   92.69  &   96.02  \\
\bottomrule

\end{tabular}
\caption{ Performance and F1-score of downstream task using sentence representation of \textsc{Bert} with and without proposed features.}
\label{tab:spreader}
\end{table}

\subsection{Fake News Spreader Classifier}
\label{sebseq::classifier}
In the next step we examine if adding the proposed features to the representation of a text will give us a better fake news spreader classifier as our downstream task. For this section, we use Bidirectional Encoder Representations from Transformers (\textsc{Bert})~\cite{devlin2018bert} as our model for sentence embedding. Each Twitter user's timeline data were represented by a vector $H_B$ by \textsc{Bert} and their motivational factors were represented by a vector $H_f$. We examined the performance of \textsc{Bert} as our baseline classification model by only using $H_B$. For the Bert+Features model the vectors are concatenated together, $H=\text{Concat}(H_B, H_f)$, and were passed through a feed-forward neural network to balance the effect of the two vectors~\cite{karami2020let}. The performance of the two models is presented in Table \ref{tab:spreader}. This experiment shows that by adding the proposed features to context representation, we can add more value to the representation of the data for classifying fake news versus real news spreaders. We give an illustrative visualization of the embeddings of 1000 random samples of Politifact and GossipCop Datasets by both \textsc{Bert} and \textsc{Bert}+Features with the t-SNE tool~\cite{van2008visualizing} (Figure~\ref{fig:tsne}). We observe that \textsc{Bert}+Features can learn more discriminative embeddings.

\section{Related Work}

Fake news detection has gained increased research attention in recent years. 
Fake news detection methods generally focus on using \textit{news content} or \textit{social context} features~\cite{shu2017fake}. 
In news content-based approaches, features are extracted linguistically or visually. Linguistic-based features capture textual information that appears in fake news content~\cite{potthast2017stylometric}, such as lexical and syntactic features. Visual-based features are used to identify fake images~\cite{gupta2013faking} that are intentionally created. 
For social context-based approaches, the features include user-based, post-based, and network-based. User-based features are extracted from user profiles~\cite{shuunderstanding} to measure their characteristics and credibility~\cite{castillo2011information}. Post-based features represent users' social responses in terms of stance~\cite{jin2016news}, topics~\cite{ma2015detect}, or credibility~\cite{castillo2011information}. Network-based features are extracted by constructing specific networks, such as diffusion networks~\cite{kwon2013prominent}.

Recent advancements utilize deep learning approaches to learn representations from the text, image, and network data for fake news detection. For example, convolution neural networks have been utilized to learn textual information from news content~\cite{wang2017liar}, and recurrent neural networks have been adopted to learn temporal representations from the propagation of fake news~\cite{ruchansky2017csi,liu2018early}. Shu \textit{et al.} utilize a network embedding approach to model the relations of the publisher, news, and users to detect fake news~\cite{shu2019beyond}. Deep geographic models such as graph convolution neural networks have been exploited to model the propagation networks as well as to detect fake news~\cite{monti2019fake}.

On the other hand, the task of profiling users has long been studied in both formal and informal languages~\cite{argamon2003gender,schwartz2013personality}. Several explicit (from metadata) and implicit cues can be used for user profiling such as user identity linkage, age, and gender~\cite{shuunderstanding}. Moreover, in recent research, factors reported from disciplines such as psychology and behavioral studies helped in profiling users as well. For instance, Giachanou et al.~\cite{giachanou2019leveraging} combined emotional signals with a Long Short Term Memory (LSTM) neural network to detect credible versus non-credible claims. Furthermore, Cardaioli et al.~\cite{cardaioli2020fake} utilized behavioral-based features such as Big Five personality and stylometric features for discriminating a fake news spreader.


\section{Conclusion and Future Work}

In this work, we investigated whether the psychological features that are observed in users who spread fake news in behavioral studies on human subjects are also visible in social media users who spread fake news on social media. Towards this goal, we introduced five categories of features based on psychological theories that can be quantified for social media users. Based on our observations on two real-world datasets, we observed that (i)~social media users who spread fake news are significantly different in terms of the majority of these features and (ii)~these features have predictive power in the detection new and unobserved fake news spreaders.

This study is a first step towards understanding users who are exposed to fake news. We wish to continue this work by looking into what other psychological features associated with individual differences such as personality traits may be related to fake news susceptibility~\cite{larson2019applying} as well as providing stronger proxy for each feature. Moreover, we plan to study the application of the psychological features in improving the performance of state-of-the-art fake news detection algorithms along with providing explanations for such algorithms~\cite{moraffah2020causal}. 
\section*{Acknowledgements}
The authors would like to thank Dr. H. Russell Bernard (ASU) and Tyler Black (ASU) for their comments on the manuscript. This material is, in part, based upon works supported by Parallax Advanced Research Corporation (11076-ASU) and ONR (N00014-21-1-4002).
\balance
\bibliographystyle{ACM-Reference-Format}
\bibliography{sample-base}
\end{document}